\documentclass[aip,apl,reprint,amsmath,amssymb,floatfix]{revtex4-1}%
\usepackage{epsf}
\usepackage{graphicx}
\usepackage{amsfonts}
\usepackage{dcolumn}%
\usepackage{color,ulem}
\usepackage{bm}

\begin{document}

\title{First-principles study of the spin-mixing conductance in Pt/Ni$_{81}$Fe$_{19}$ junctions }
\author{Qinfang Zhang}
\email{q.zhang@riken.jp}
\author{Shin-ichi Hikino}
\affiliation{Computational Condensed Matter Physics Laboratory, RIKEN ASI, Wako, Saitama 351-0198, Japan \\}
\affiliation{CREST, Japan Science and Technology Agency, Kawaguchi, Saitama 332-0012, Japan \\}
\author{Seiji Yunoki}
\affiliation{Computational Condensed Matter Physics Laboratory, RIKEN ASI, Wako, Saitama 351-0198, Japan \\}
\affiliation{CREST, Japan Science and Technology Agency, Kawaguchi, Saitama 332-0012, Japan \\}
\affiliation{Computational Materials Science Research Team, RIKEN AICS, Kobe, Hyogo 650-0047, Japan \\}

\date{\today}

\begin{abstract} 
Based on the spin-pumping theory and first-principles calculations, the spin-mixing conductance (SMC) is theoretically studied for Pt/Permalloy (Ni$_{81}$Fe$_{19}$, Py) junctions.
We evaluate the SMC for ideally clean Pt/Py junctions and examine the effects of interface randomness. We find that the SMC is generally enhanced in the presence of interface roughness as compared to
the ideally clean junctions.
Our estimated SMC is in good quantitative agreement with the recent experiment for Pt/Py junctions.
We propose possible routes to increase the SMC in Pt/Py junctions by depositing a foreign magnetic metal
layer in Pt, offering guidelines for designing the future spintronic devices.\end{abstract}

\maketitle

Spintronics utilizes the electron spin degree of freedom for device applications such as data storage, 
non-evaporate memory, and high speed processing, which exceed the conventional electronics.~\cite{zutic@rmp04} 
These spintronic devices are controlled by spin current, 
and thus efficiently creating the spin current is one of the primary issues.~\cite{maekawa}

One of the standard ways to generate the spin current is the spin-pumping through a ferromagnetic metal/normal metal
(F/N) junction where the spin current is pumped out from the F into the N by ferromagnetic resonance 
in the F.~\cite{tserkovnyak@prl02} 
The spin current induced by the spin-pumping is proportional to the spin-mixing 
conductance (SMC),~\cite{tserkovnyak@prl02, zwierzycki@prb05, kardasz@prb10} and 
thus the large SMC is required for highly efficient F/N junctions as a spin current generator.

The large SMC is also related to the enhanced Gilbert magnetic damping in the F.~\cite{kardasz@prb10,ghosh@apl11} 
For instance, 
the critical current density of the current induced magnetization reversal is proportional to 
the Gilbert damping constant, and the large Gilbert damping is suitable for fast switching of magnetization reversal.~\cite{sun} 
Thus, the SMC is also an essential parameter to design high speed and low power consumption spintronic devices. 
In spite of its importance, the SMC is usually employed as a fitting parameter of experimental results.~\cite{ando} 
Therefore, it is highly desirable to quantitatively evaluate the SMC based on first-principles 
calculations.~\cite{zwierzycki@prb05,carva@prb07}

In this Letter, based on the spin-pumping theory~\cite{tserkovnyak@prl02} and first-principles calculations, we 
investigate the SMC in Pt/Permalloy (Ni$_{81}$Fe$_{19}$, Py) junctions. We numerically evaluate the SMC for the 
ideally clean junctions with three different crystalline orientations, and discuss quantitatively the effects of interface 
randomness. 
We find that the interface roughness generally enhances the SMC as compared to the ideally clean junctions. 
Our estimated SMC is found to be in good agreement with the recent experiment.~\cite{ando}
We also discuss possible routes to increase the SMC in Pt/Py based junctions by depositing an additional buffer layer in Pt or Py. 
Our results indicate that a key ingredient to increase the SMC is to deposit in Pt a magnetic layer such as a Fe layer.

Here, the SMC is evaluated based on the spin-pumping theory and using first-principles calculations. 
In the spin-pumping theory,~\cite{tserkovnyak@prl02} the SMC is given by 
$G_{r}^{\uparrow \downarrow} = S^{-1}\sum_{m,n}(\delta_{mn} - r_{mn}^{\uparrow } r_{mn}^{\downarrow \star })$, 
where $S$ is the contact area, $m$ and $n$ denote scattered electronic states at the Fermi energy of the nonmagnetic 
Pt lead, and $r_{mn}^{\uparrow (\downarrow)}$ is the reflection matrix at the interface 
for up (down) electrons.~\cite{note1} 
Taking into account the realistic electronic structures in the Pt/Py junctions by first-principles calculations, 
the reflection coefficients $r_{mn}^{\uparrow (\downarrow)}$ are estimated.

\begin{figure}
\includegraphics[scale=1.0]{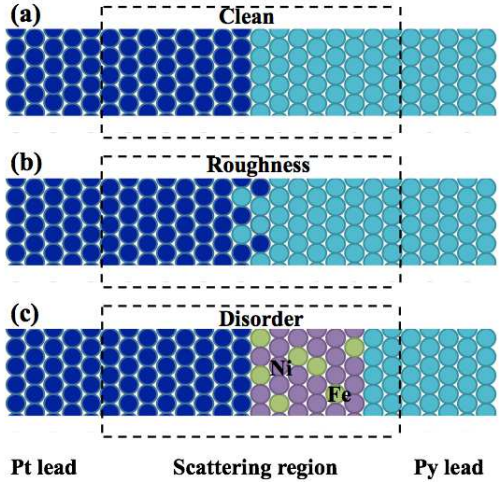}
\caption {
(Color online) Schematic figures of Pt/Py junctions with Pt$\vert$S$\vert$Py structure.  Interface structures are characterized 
in a scattering region (S) (indicated by dotted lines) for (a) ideally clean case with S=Pt(8)Py(8), and (b) interfacially rough case 
with S=Pt(7)[Pt$_{1-x}$Py$_x$](1)[Pt$_{x}$Py$_{1-x}$](1)Py(7). 
The number in parentheses denotes the number of atomic layers in the S. 
(c) The disordered alloy character of Py is modeled by randomly locating Ni and Fe atoms 
with Ni$_{81}$Fe$_{19}$ ratio in the S, where two layers of Py are also included at the right end of the S. 
} 
\label{schematic}
\end{figure}

We first perform a self-consistent tight-binding linearized muffin-tin orbital (TB-LMTO) 
calculation~\cite{andersen@prb86} with atomic sphere approximation for a system consisting 
of semi-infinite Pt and Py leads, and a scattering region (S) which includes eight layers of Pt and Py 
for the clean interface,~\cite{note5} as schematically shown in Fig.~\ref{schematic} (a). Hereafter this system is denoted by 
Pt$\vert$S$\vert$Py for short, and thus the ideally clean junction is Pt$\vert$Pt(8)Py(8)$\vert$Py 
with the number in parentheses indicating the number of atomic layers. 
The charge and spin densities of Py are treated using virtual crystal approximation (VCA).~\cite{vca-cpa} 
We use \textit{spdf} muffin-tin orbital basis set to solve the Schr$\rm{\ddot o}$dinger equation. 
The exchange-correlation potential is parameterized 
according to von Barth and Hedin.~\cite{barth@jpc72}  The experimental lattice constant (a=3.55\AA{}) of Py 
with {\rm f.c.c.} structure~\cite{lattice} is chosen for both Pt and Py 
leads for simplicity.~\cite{mismatch} 
Once the atomic potential in the S is determined self-consistently, 
the transmission 
matrix $r_{mn}^{\uparrow (\downarrow)}$  is calculated using a TB-MTO implementation of the wave function matching 
scheme and the conductance 
is evaluated using Landauer-B\"{u}ttiker formula.~\cite{xia} This method has been successfully applied for a number of spin 
transport calculations including the spin dependent interface resistances and the thermal spin transfer torque in 
magnetoelectronic devices.~\cite{xia,hatami}

To study the effects of interface randomness, we also consider randomly rough interface. 
The interface roughness is treated by substituting 
a composite Pt$_{1-x}$Py$_x$ and Pt$_{x}$Py$_{1-x}$ layer for the first Pt and Py layer at the interface, respectively, 
as illustrated in Fig.~\ref{schematic} (b).~\cite{note2} 
Thus, the S is described by Pt(7)[Pt$_{1-x}$Py$_x$](1)[Pt$_{x}$Py$_{1-x}$](1)Py(7). 
The atomic potential 
in the S is calculated using the coherent potential 
approximation (CPA).~\cite{vca-cpa} 
The concentration $x$ is varied from 0 to 0.5 with $x=0$ corresponding to the clean interface. 
For each $x$, we generate at least 16 different random configurations of atomic positions in the interfacially 
rough layers, and the SMC is estimated by averaging the results for each random configuration. 
In addition, to validate the VCA treatment for Py, we study the disordered alloy effects of Py by explicitly 
including disordered Ni and Fe atomic positions with Ni$_{81}$Fe$_{19}$ ratio in the S [Fig.~\ref{schematic} (c)], 
where Ni and Fe in the S are treated by CPA.

\begin{figure}
\includegraphics[scale=0.4]{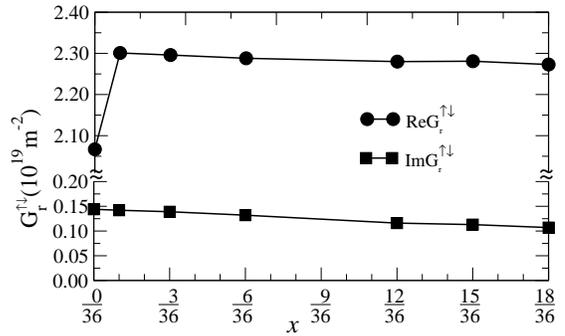}
\caption { 
The SMC in the Pt/Py junctions with the interface roughness, corresponding to Fig.~\ref{schematic} (b). 
The S in Pt$\vert$S$\vert$Py is described by Pt(7)[Pt$_{1-x}$Py$_x$](1)[Pt$_{x}$Py$_{1-x}$](1)Py(7). 
The lateral size of the supercell used~\cite{note2} is $6\times6$ 
and the crystalline orientation is along [111] direction.
} 
\label{erro}
\end{figure}

Let us first study $x$\textcolor{blue}{-}dependence on the SMC in the Pt/Py junctions with $[111]$ crystalline orientation. 
The obtained results are shown in Fig.~\ref{erro}. It is found that i) the imaginary part of the SMC is approximately 
one order of magnitude smaller than the real part of the SMC, and ii) the real part of the SMC is larger for the rough 
interfaces (finite $x$) than for the clean interface ($x=0$). Moreover, we can see in Fig.~\ref{erro} that the SMC is 
rather insensitive to $x$ as long as $x$ is finite, and thus we choose $x=0.5$ for the Pt/Py junctions with the interface 
roughness studied below.

The results for the SMC with other crystalline orientations, including the one shown above, are summarized in Table~\ref{tab:conductance}.  
We can see in the table that  the imaginary part of the SMC (Im~$G^{\uparrow\downarrow}_{r}$) is about 7 \% or 
less smaller than the real part (Re~$G^{\uparrow\downarrow}_{r}$), which is similar to the previous studies on different F/N 
junctions.~\cite{xia}
It is also noticed that Re~$G^{\uparrow\downarrow}_{r}$ is close to the Sharvin conductance of 
Pt ($2.553 \times 10^{19} \rm m^{-2}$).~\cite{zwierzycki@prb05}
Moreover, we find that Re~$G^{\uparrow\downarrow}_{r}$ is generally larger for the junctions with the interface roughness than for 
the ideally clean interface, indicating that the interface roughness plays an important role to achieve a large SMC. 
Our estimated SMC is in good quantitative agreement with the recent experiment for Pt/Py junctions where the SMC is observed 
to be about $2.3\times10^{19} {\rm m^{-2}}$.~\cite{ando} 
Finally, in Table~\ref{tab:conductance}, also shown is the SMC for the disordered Py alloy [Fig.~\ref{schematic} (c)], where 
we find that the SMC is almost the same for the clean junction illustrated in Fig.~\ref{schematic} (a), justifying VCA for Py alloy.

\begin{table}
\caption{\label{tab:conductance}
The SMC $G^{\uparrow\downarrow}_{r}$ (in units of 10$^{19} {\rm m^{-2}}$) in the Pt/Py junctions with three different 
crystalline orientations. The interface roughness is simulated with $x=0.5$. 
For comparison, the SMC for the disordered Py alloy [Fig.~\ref{schematic} (c)] is also listed. 
}
\resizebox{6.50cm}{!}{
\begin{tabular}{cccc}
\hline
\hline
Orientation&Interface     &${\rm Re}G^{\uparrow\downarrow}_{r}$&${\rm Im}G^{\uparrow\downarrow}_{r}$\\
\hline
$[001] $  &clean            &2.174  &0.123 \\
$[001]$   &roughness  &2.347  &0.185 \\
$[110] $  &clean           &2.354  &0.128 \\
$[110] $  &roughness  &2.387  &0.114 \\
$[111] $  &clean           &2.066  &0.141 \\
$[111] $  &roughness  &2.273  &0.106 \\
$[111] $  &disorder      &1.996  &0.055 \\
\hline
\hline
\end{tabular}
}
\end{table}

\begin{figure}
\includegraphics[scale=0.8]{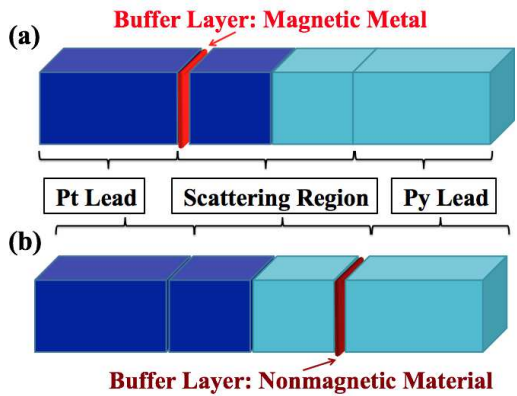}
\caption {
(Color online) Schematic illustrations of the Pt/Py junctions with (a) a magnetic atomic layer and (b) 
a nonmagnetic atomic layer deposited as a buffer layer in Pt and Py layers, respectively. 
} 
\label{buffer}
\end{figure}

Let us now discuss how to increase the SMC in the Pt/Py based junctions. 
The SMC is essentially determined by the interface property since this quantity depends on the reflection coefficient at the interface. 
Therefore, we expect its value to vary by controlling the interface structure. 
A possible way to obtain a larger SMC is to deposit a buffer layer such as a transition metal (nonmagnetic atomic) layer 
for a Pt (Py) layer, 
as schematically 
shown in Fig.~\ref{buffer}. We consider seven different buffer layers and the results are listed in Table~\ref{tab:enhance}. 
We find that the SMC generally increases when a ferromagnetic layer such as Py and Fe, whose magnetic moment is also shown in 
Table~\ref{tab:enhance}, is deposited in the Pt layers. 
Especially, ${\rm Re}~G^{\uparrow\downarrow}_{r}$ becomes largest 
(${\rm Re}~G_{r}^{\uparrow \downarrow }\approx 2.5 \times 10^{19} {\rm m^{-2}}$)
when two layers of Fe are deposited. 
Instead, ${\rm Re}~G^{\uparrow\downarrow}_{r}$ remains almost the same when nonmagnetic metal such as Pt is deposited 
in the Py layers. 
Moreover, we can model a tunneling junction by depositing
an insulating layer (I) in the Py layers,~\cite{note4} and the results are also shown in Table~\ref{tab:enhance}. 
It is interesting to notice that the SMC in the tunnel junction remain large 
(${\rm Re}~G_{r}^{\uparrow \downarrow }\approx 2.4 \times 10^{19} {\rm m^{-2}}$) even though the 
conductance is vanishingly small. 
A general tendency that we find empirically through this study is that the SMC increases when a magnetic layer is deposited 
in the nonmagnetic Pt layers in the vicinity of the interface, which we believe can provide guidelines for designing the future 
spintronic devices.

\begin{table}
\caption{\label{tab:enhance}
The SMC (in units of 10$^{19} \rm m^{-2}$) in the Pt/Py junctions (Pt$\vert$S$\vert$Py) with the clean interface. 
The crystalline orientation is along [111] direction.  
The conductance ($G$ in units of 10$^{19}$ $\rm{m^{-2}}$) and 
the magnetic moment ($M$) per layer in the deposited buffer layer(s) 
(normalized by the Bohr magneton $\mu_{\rm B}$) are also listed. 
}
\resizebox{8cm}{!}{
\begin{tabular}{ccccc}
\hline
\hline
Pt$\vert$S$\vert$Py   &${\rm Re}G^{\uparrow\downarrow}_{r}$&${\rm Im}G^{\uparrow\downarrow}_{r}$&$G$   
&  $M/\mu_{B}$\\
\hline
Pt$\vert$Pt(8)Py(8)$\vert$Py      & 2.066& 0.141&2.634     & \\   
Pt$\vert$Py(1)Pt(7)Py(8)$\vert$Py &2.215&0.261& 2.382&0.81\\ 
Pt$\vert$Fe(1)Pt(7)Py(8)$\vert$Py&2.230&0.374&1.861&2.09\\
Pt$\vert$Fe(2)Pt(6)Py(8)$\vert$Py&2.520&0.228&1.859&2.03\\
Pt$\vert$Ni(1)Pt(7)Py(8)$\vert$Py&1.900&0.074&2.442&0.01\\
Pt$\vert$Pt(8)Py(7)Pt(1)$\vert$Py&2.068&0.122&2.471&0.19\\
Pt$\vert$Py(1)Pt(7)Py(7)Pt(1)$\vert$Py&2.179&0.301&2.212&0.9,0.2\\
Pt$\vert$Pt(8)Py(7)I(1)$\vert$Py&2.356&0.187&0.343&0\\
\hline
\hline
\end{tabular}
}
\end{table}

In summary, we have studied the SMC in the Pt/Py junctions taking into account the realistic electronic structures 
by using first-principles calculations. 
We have evaluated the SMC for the ideally clean junctions 
and examined the influences of the interface roughness. 
We have found that the imaginary part of the SMC is approximately one order of magnitude smaller 
that the real part of the SMC, and that, 
as compared to the ideally clean junctions, the interface roughness generally enhances the real part of the SMC. 
We have furthermore discussed possible routes to increase the SMC in Pt/Py based junctions by 
depositing additional 
buffer layers in Pt or Py layers. Our results indicate that the SMC increases when a magnetic layer is deposited 
in the nonmagnetic Pt layers.  
These results offer valuable guidelines for designing the future spintronic devices.

The authors thank S. Maekawa and J. Ieda for variable comments. 
A part of the calculation has been performed using RIKEN Integrated Cluster of Clusters (RICC).

{}


\begin{thebibliography}{}
\bibitem{note1}
The general form of the SMC contains $G_{r}^{\uparrow \downarrow }$ and $G_{t}^{\uparrow \downarrow }$, 
where  $G_{t}^{\uparrow \downarrow }=S^{-1}\sum_{mn}t_{mn}^{\uparrow }t_{mn}^{\downarrow \star }$ and  
$t_{mn}^{\uparrow (\downarrow)}$ is the transmission matrix for up (down) electrons. 
However, in metallic junctions, $|G_{t}^{\uparrow \downarrow }|$ sharply decreases to zero with the thickness 
of magnetic layers,~\cite{zwierzycki@prb05} and thus $G_{t}^{\uparrow \downarrow }$  can be safely neglected 
for thicker junctions.
\bibitem{note5} 
We set eight layers of Pt and Py in the S to ensure that the electronic structures away from
the interface are similar to the bulk.
\bibitem{mismatch} We have checked the effect of the lattice constant mismatch for Pt and Py by using the lateral supercell sizes of 
Pt ($9\times9$) and Py ($10\times10$), and found that the real part of SMC generally decreases at most about 10 \%. However, 
the main conclusions in this paper are unaffected.
\bibitem{note2}

We have checked the lateral supercell size ($N$) dependence of the SMC up to $N=14\times14$, and found that  
the SMC is already converged for $N=6\times6$. Therefore, 
we have adopted $N=6\times6$ for calculations of the SMC presented here.
\bibitem{note4}
A vacuum barrier is adopted as an insulating buffer layer where 
we use empty spheres to populate the vaccine sites.
\end{thebibliography}
\end{document}